\documentclass[manuscript,nonacm]{acmart}

\AtBeginDocument{%
  }

\usepackage{array, caption, tabularx,  ragged2e}
\usepackage{longtable}
\usepackage{enumitem}
\usepackage{multirow}
\newcolumntype{L}{X}
\newcolumntype{M}{>{\hsize=.5\hsize}X}
\newcolumntype{S}{>{\hsize=.25\hsize}X}

\definecolor{Red}{rgb}{1,0,0}
\definecolor{Green}{rgb}{0,1,0}
\definecolor{Blue}{rgb}{0,0,1}

\begin{document}

\title{Human–AI Writing as Distributed Cognition: Reassessing Collaborative Writing Theories in the Age of LLMs}

\author{Daisuke Yukita}
\email{d.yukita@uq.edu.au}
\orcid{0009-0003-9050-800X}
\affiliation{%
  \institution{The University Of Queensland}
  \city{Brisbane}
  \country{Australia}
}

\author{Tim Miller}
\orcid{0000-0003-4908-6063}
\affiliation{%
  \institution{The University Of Queensland}
  \city{Brisbane}
  \country{Australia}
}
\email{timothy.miller@uq.edu.au}

\author{Joel Mackenzie}
\orcid{0000-0001-7992-4633}
\affiliation{%
  \institution{The University Of Queensland}
  \city{Brisbane}
  \country{Australia}
}
\email{joel.mackenzie@uq.edu.au}

\renewcommand{\shortauthors}{Yukita et al.}

\begin{abstract}
  In this paper, we conduct a structured synthesis of existing theories and frameworks on human-human collaborative writing to assess how co-writing with AI reshapes writing in knowledge work settings, and draw seven insights along with design implications for human-AI collaborative writing tools. Our main argument is that, as we delegate more writing to AI, our cognitive process shifts from the traditional plan/write/review process to a plan/delegate/review process, reshaping the distribution of the cognitive labor between human writers and AI writing tools. To ensure that humans cognitively stay in the loop, we discuss a ``prototyping'' approach, where the writing tool scaffolds a healthier distribution of cognition through rapid, small-scale iterations that keep writers mentally in the loop even when generation is delegated to AI. We aim to bring theoretical grounding and practical guidance to the interaction designs of human-AI collaborative writing, with the goal of balancing automation with cognitive integrity.
\end{abstract}


\begin{teaserfigure}
  \includegraphics[width=\textwidth]{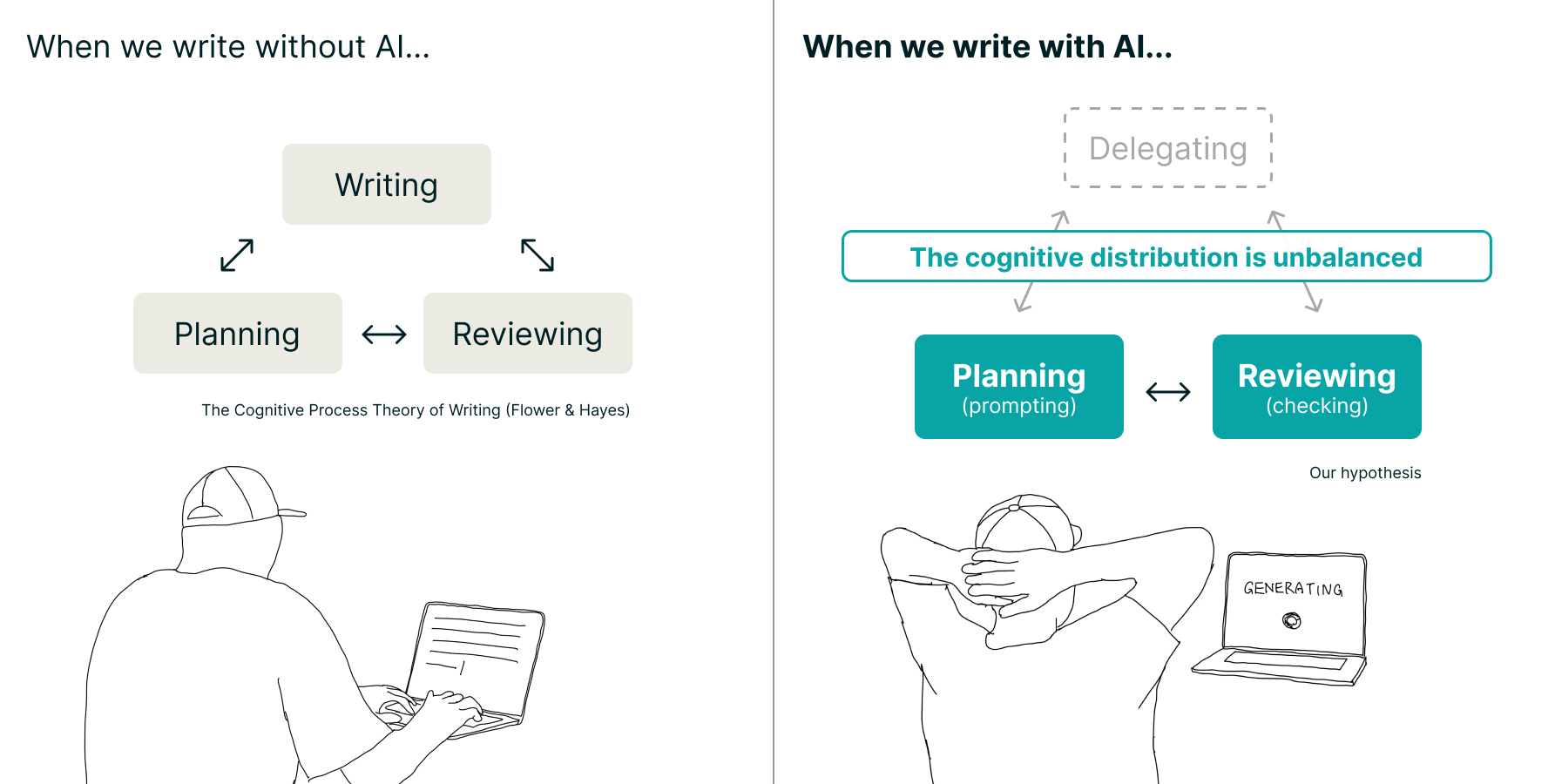}
  \caption{How we see AI to affect our cognitive writing process.}
  \Description{A sketch comparing the cognitive process of writing with and without AI. Traditionally our cognitive writing process was described as the iterative and non-linear cycle of plan/write/review, but the delegation of composition to AI skews the cognitive labor due to offloading.}
  \label{fig:teaser}
\end{teaserfigure}

\maketitle
\pagestyle{empty}

\section{Introduction}
When humans made the shift from writing on paper to typing on word-processors in the 1980s, studies found that writers engaged in less conceptual planning, reread texts more extensively, and recalled their own text less~\cite{Haas2013-av}. How then, will writing with Large Language Models (LLMs) affect the way that we write, and what negative impacts should we aim to mitigate in designing LLM-based writing tools? As LLMs increasingly become our daily co-writers, and as their capabilities in natural language generation become increasingly indistinguishable from human writing~\cite{Hamalainen2023-gc, Clark2021-vx, Zhu2024-ht}, we see a need to reassess foundational theories and frameworks on collaborative writing, and see what they can inform us about the human-AI paradigm. To reflect this paradigm shift, new frameworks and taxonomies of collaborative writing are emerging, from domain-specific taxonomy development~\cite{Lee2024-dz}, the ``AI ghostwriter effect''~\cite{Draxler2024-es} (the finding that human writers tend to attribute ownership of AI generated text to AI but refrain from publicly disclosing it), analysing AI-assisted writing patterns using Hidden Markov Model and pattern analysis~\cite{Nguyen2024-nn}, to exploring a non-linear framework to go beyond the linear human-only/human-AI/AI-only dimension of collaboration modes~\cite{Hutson2025-fp}. While many of these proposals build on a partial aspect of existing theories and frameworks on collaborative writing, we believe that this must be approached holistically, synthesizing what past studies have taught us about the mechanisms of collaborative writing. Naturally, legacy theories and frameworks on collaborative writing have focused on humans writing with humans. To what degree do these understandings of human-human collaborative writing apply to the human-AI paradigm? We believe now is the moment to look back to the diverse literature in collaborative writing, and see what we can take from them to the emerging human-AI mode.

Hence we arrive to our research question: \textbf{What insights can we draw from theories/frameworks on human-human collaborative writing to apply to the design of emerging human-AI writing tools? What aspects of those theories/frameworks are less applicable for human-AI collaboration?}

In this paper, we conduct a structured synthesis of legacy theories and frameworks on collaborative writing, particularly through the lens of distributed cognition, and analyze what we see to be relevant to the human-AI paradigm as well as what only applies to the human-human paradigm. Our main argument is that, when the human's role shifts from a hands-on {\emph{writer}} to an editorial {\emph{decision maker}}, the cognitive process of writing~\cite{Flower1981-ey} shifts from the traditional \textit{plan/write/review} process to \textit{plan/\textbf{delegate}/review} (Figure~\ref{fig:teaser}), which externalizes the composing act and interrupts cognitive continuity due to excessive offloading. In response to this, we discuss opportunities for AI tools to scaffold a healthier distribution of cognitive labor, such as through rapid, small-scale iterations of text that keep writers mentally in the loop even when generation is delegated to AI, supporting human writers to revise AI generated text faster, and helping them connect that to the next iteration of text generation. Coherence support between human and AI writing is an ongoing challenge, but the strong semantic capabilities of LLMs can provide real-time analyses that were previously difficult with traditional methods. We assert that teamwork-related factors such as group awareness, consensus building and co-authorship, which have been central in human-human collaborative writing studies, will be less applicable in the human-AI paradigm due to excessive anthropomorphism, but it is equally the AI co-writer's role to facilitate such factors to offload teamwork-related mental workload. Finally, be it human-written or AI-written, writing will remain to be socially constructed by readers, as long as humans continue to be the primary readers of the documents we create (which admittedly may not always be the case for all documents). This suggests that the consideration of reader profiles is inseparable from the text generation process. 

We provide practical design implications and example interface sketches to provide concrete design directions. To the best of our knowledge, this is the first work to holistically reassess legacy collaborative writing theories and frameworks in the context of the LLM era.

The main contributions of this work are as follows:
\begin{itemize}
        \item we define the limitations of existing human-human collaborative writing theories and frameworks when applied to the human-AI paradigm;
        \item we propose an updated cognitive model of collaborative writing with AI that builds on Flower and Hayes'~\cite{Flower1981-ey} cognitive process of writing and highlights imbalances in the distribution of cognitive labor; and
    	\item we provide actionable design implications for future human-AI writing tools based on what we can apply from existing collaborative writing theories.
\end{itemize}

Ultimately work aims to provide initial footing for theoretical grounding to rapidly evolving AI writing studies, allowing more researchers to build and connect with what we already know about how humans seek to write with others.

\subsection{Study scope}
\label{scope}
Writing is a highly complex act of cognition and creativity, and is the bread and butter for some professions like professional writers or researchers. For these professions, it is unlikely that writers will want to consider delegating their core competency to AI~\cite{Zulu2026-kq}, nor should they necessarily be encouraged to do so from an accountability perspective~\cite{Stokel-Walker2023-ry}. For this reason, the scope of this paper is limited to the pragmatic support of writing tasks for knowledge workers~\cite{Davenport2005-wx} in organizational workplace settings. A prototypical example of such case would be drafting emails with the help of LLMs, clinicians drafting discharge summaries, course coordinators drafting a response to student questions on online forums, meeting participants drafting a summary to send out to their team, or academics crafting grant proposals. For the same reason, we do not consider creative or fictional writing \cite{Yang2022-ff, Calderwood2020-wg}, nor do we focus on the deskilling concerns of human writing \cite{Kosmyna2025-mp}, even though we agree with the risks of this. Building on the views of Naikar et al.~\cite{Naikar2025-sq} and Schneiderman~\cite{Shneiderman2022-ke} we take the stance that LLMs are a tool to help us get the job done in the most effective manner, and avoid framing the AI as a ``team mate'' that shares writing goals and motivations.

\section{Background}
In this section, we revisit what the field of collaborative writing has taught us about the mechanisms of human-human writing, and reframe this from a distributed cognition perspective. This reframing builds the core foundation of our claim that AI tools need to scaffold the cognitive delegation to make sure humans \textbf{stay}-in-the-loop.

\subsection{Collaborative Writing}
While writing was once thought to be a solitary act {\cite{foucault69}}, there was a gradual realization that, in many cases, humans in fact write together. This is often the case in work settings, such as engineers articulating project requirements, psychologists drafting brochures for stress management programs, or technical writers creating construction manuals \cite{Ede2006-wh}. To reflect this, the widely adopted theory of an individual's cognitive writing process by Flower and Hayes~\cite{Flower1981-ey} was updated by Hayes 15 years later~\cite{Hayes1996-ho}, emphasizing to the ``social'' aspect of writing and incorporating collaborators as well as readers into their framework.
Ede and Lunsford~\cite{Lunsford1986-jo} were the among the first to initiate collaborative writing as an academic research topic in 1986, reporting that 87 percent of respondents from six major professions sometimes wrote as a team or a group. Through further analysis they found that there were two distinct modes in collaborative writing:  the ``hierarchical'' mode where one person has control and requires structure and division of labor, and the ``dialogic'' mode where all participants have equal input with shared authorship, which requires consensus and compromise~\cite{Ede2006-wh}.
However, they also quickly realized that there was a significant lack of shared consensus on what collaborative writing {\emph{is}}. As Allen et al.~\cite{Allen1987-zg} assert: ``very little detail is known about collaborative writing processes in general.'' Due to the highly complex nature of document structures as well as the strong connection between writing and subjective emotions, collaborative writing is perhaps one of the most difficult forms of collaboration. As Farkas explains (in Lay and Karis~\cite{LayUnknown-jh}), there may not even be a common language for collaborative writing beyond a certain level.
In light of this, Lowry et al.~\cite{Lowry2004-me} established the taxonomy of collaborative writing, outlining not just the various roles but also the writing strategies that groups adopt when writing together. As the role of computers and software are becoming increasingly important in writing, this taxonomy is getting updated to this day, including Collage~\cite{Buschek2024-ka}, CoCo matrix~\cite{Wan2024-rd}, Overall Writing Effectiveness~\cite{Wilbers2024-kl}, neurodiversity-inspired non-linear framework~\cite{Hutson2025-fp}, taxonomy of ``text functions''~\cite{Larsen-Ledet2020-nz} to name a few.
While one may question as to what extent writing with AI can be considered ``collaborative writing'', when a human tasks an LLM to write up a draft of a document the human-AI pair is working ``toward coordinated consensus that is reflected in a document that is written by one of the team members''~\cite{Lowry2004-me}, which we see to qualify as a form of collaborative writing. It is also worth noting here that, according to Posner and Baecker~\cite{Posner1992-ff}, the single writer strategy was the most common strategy adopted, even in human-only groups. If so, it seems viable to consider writing with AI to be a legitimate form of collaborative writing. Be it with humans or with AI, writing is hardly a solitary act: it happens through the participation and interaction with others~\cite{Ede2006-wh}.

\subsection{Distributed cognition}
\label{distributed-cognition}
While the Flower \& Hayes' ``plan/write/review'' model~\cite{Flower1981-ey} has been used as the foundational theory of many collaborative writing studies, we reframe this through the distributed cognition lens to take it from human-human collaboration, to the current human-AI paradigm~\cite{Grinschgl2022-qk}. The theory of distributed cognition sees human cognition to extend beyond our individual minds, as something that is formed through the interaction between other minds or artifacts~\cite{Hollan2000-lr}. More specifically, there are three aspects to distributed cognition: (1) socially distributed cognition (cognitive processes that are distributed with other humans, such as group work); (2) temporally distributed cognition (cognitive processes that are distributed through time, such as past usage); (3) and externalised cognition (cognitive processes that are distributed with other physical artifacts, such as computers or diaries)~\cite{Hollan2000-lr}. In other words, in the distributed cognition construct, cognition emerges across people, time and tools. Collaborative writing with AI can be categorized particularly as externalised cognition, where parts of the writing process are offloaded to an AI tool, and the writing itself is jointly enacted between human writers and AI. While the notion of distributed cognition in collaborative writing in itself is not new~\cite{Ross1995-og}, developing a shared understanding of text with other human writers is different in nature from developing a shared understanding with an AI tool.

By reframing the cognitive writing process as a distributed system, it allows us to analyze the current challenges of co-writing with AI with more structural clarity. For instance, Kosmyna et al.~\cite{Kosmyna2025-mp} found that when students were asked to write essays just with ChatGPT, they not only struggled to quote phrases from the essay they wrote, but many failed to even recall the topic they chose to write about when they were called back a few weeks after the writing task. They warn that writing with LLMs could lead to decreased cognitive agency, which would further lead to reduced error-monitoring~\cite{Kosmyna2025-mp}. From a distributed cognition standpoint, this could be explained as an imbalance of cognitive distribution: human writers are excessively offloading the writing task to AI tools, losing the ability to stay cognitively engaged in the content. The Flower/Hayes model~\cite{Flower1981-ey} has taught us that writing processes are inherently non-linear and iterative -- we think as we write, write as we review, and so forth. Delegating the writing to generative AI externalizes the composing act entirely, which disrupts the cognitive continuity for the human writer.

It is important to note here that we do not necessarily see delegation itself as a problem, because for any task, offloading comes with both benefits and risks~\cite{Grinschgl2022-qk}. With writing specifically, however, the physical act of composing text is a central mechanism of our cognitive labor \cite{Mueller2014-zq}, and taking that away leads to detrimental effects for our cognitive process. Hence we argue that AI writing tools must support the optimal distribution of cognitive labor so that humans remain cognitively coupled, even amidst AI delegation.

\section{Survey Method}

\begin{figure}[!t]
  \centering
  \includegraphics[scale=0.4]{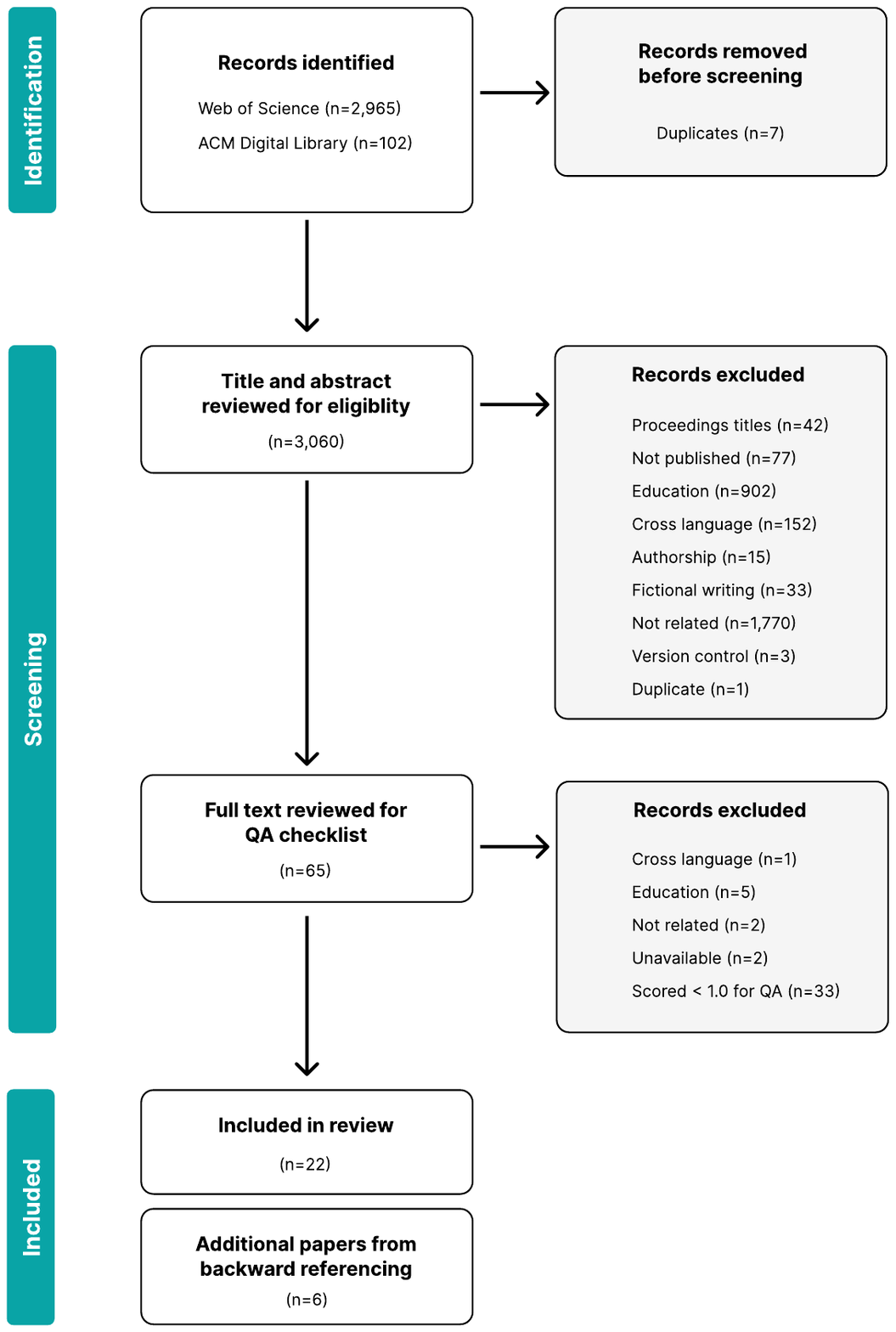}
  \caption{A PRISMA diagram for the methodology used in our review. Our initial search yielded over $3{,}000$ articles, with $28$ retained for final review.}
  \Description{A PRISMA diagram for the methodology used in our review.}
  \label{fig:prisma}
\end{figure}

Our review followed the PRISMA methodology~\cite{Moher2009-cr}. The PRISMA diagram for our review is presented in Figure~\ref{fig:prisma}. To serve our core research questions, the questions for the systematic review were as follows:

\begin{enumerate}
        \item What theories/frameworks exist for collaborative writing that can be applied to LLM interactions for practical writing tasks today, and which theories/frameworks are not relevant?
    	\item Which studies in HCI already build on these theories/frameworks when designing for AI-assisted writing?
        \item What design implications can we draw from these theories and frameworks to apply to current human-LLM interaction?
\end{enumerate}

Drawing on the principal search system suggestions by Moher et al.~\cite{Moher2009-cr}, we selected 2 databases (Web of Science and the ACM Digital Library) and applied the queries in Table~\ref{tab:queries}. It is important to note that this study aims to bridge existing collaborative writing theories/frameworks (which fall under various research fields, such as rhetoric and writing, cognitive science, and Computer-Supported Cooperative Work) with AI writing studies (which are often from HCI). For this reason, we relied on the single phrase “collaborative writing” for the Web of Science query to cast an intentionally wide net -- we wanted to gather as much existing work from outside the fields of HCI and computer science, new and old. At the same time, we intentionally added a filtering for LLM related keywords for the ACM library query, because the selection of the ACM library was primarily to address RQ2. This was not necessary for Web of Science due to its broader, non-HCI-specific scope. Given that collaborative writing has been studied for decades, we also did not limit by publication year.

\begin{table*}[t]
    \centering
    \renewcommand{\arraystretch}{2}
    \begin{tabular*}{\textwidth}{@{\extracolsep{\fill}}p{0.2\textwidth}p{0.8\textwidth}@{}}
        \textbf{Database} & \textbf{Query} \\
        \hline
        Web of Science &
        \textit{"collaborative writing"} {\sf{AND}} (\textit{"theory"} {\sf{OR}} \textit{"theories"} {\sf{OR}} \textit{"framework"} {\sf{OR}} \textit{"taxonomy"} {\sf{OR}} \textit{"taxonomies"})
        \\
        ACM Digital Library &
        \textit{"collaborative writing"} {\sf{AND}} (\textit{"theory"} {\sf{OR}} \textit{"theories"} {\sf{OR}} \textit{"framework"} {\sf{OR}} \textit{"taxonomy"} {\sf{OR}} \textit{"taxonomies"}) {\sf{AND}} (\textit{"large language models"} {\sf{OR}} \textit{"llm"} {\sf{OR}} \textit{"generative ai"} {\sf{OR}} \textit{"gen ai"})
        \\
        \hline
    \end{tabular*}
    \vspace{10pt}
    \caption{The two search queries executed in our literature review.}
    \label{tab:queries}
\end{table*}

These queries resulted in a total of 3,067 entries (2,965 from Web of Science, 102 from ACM). After removing 7 duplicates, we applied the inclusion/exclusion criteria in Appendix~\ref{appendix:a} to the 3,060 entries based on the titles and abstracts. In addition to fictional writing, rhetoric, and authorship, we also excluded studies in the educational context, such as those related to the pedagogy of collaborative writing or second language learners. This is because, in the context of knowledge work, we see such considerations to be largely irrelevant -- the primary aim for such users is to simply get the task done with less time and effort. As mentioned in Section~\ref{scope}, the scope of this paper is limited to the pragmatic support of writing tasks in organizational workplace settings. While educational approaches in these settings should in theory be of benefit, we do not see this as the primary goal for AI-writing tools.
This left us with 65 entries -- we lightly reviewed the full text, and screened with the predefined QA checklist in Appendix~\ref{appendix:a}. We applied a threshold of $1$ for the QA checklist (i.e. we only retained papers that achieved a full score of $1$), leaving us with 22 entries. Finally, we added 6 papers from backward referencing, resulting in a total of 28 papers for the full review.

Following study selection, we conducted a qualitative synthesis to identify recurring theoretical constructs and their implications for human-AI writing. For each paper we first performed data extraction, recording the collaborative writing theories/frameworks that the work introduced or built on top of, along with key observations related to writing processes, collaboration patterns and designs of AI-assisted writing tools. We then conducted an iterative, inductive thematic analysis, by progressively grouping and labeling notes based on conceptual similarity, allowing higher level insights to emerge. Rather than applying a predefined coding scheme, themes were developed through the iterative process. This process resulted in a set of theoretically grounded themes that capture the characteristics of how humans have traditionally written with others, which we will subsequently refer to as ``themes'' in this paper. Further, based on the findings from AI-assisted writing tools presented in the reviewed papers (from ACM in particular) as well as recent literature on human-AI collaboration, we examined how each theme informs or challenges existing assumptions of human-AI writing, and derived implications for the design of human-AI writing tools which we subsequently refer to as ``design implications''. It is here that we applied the distributed cognition lens, using it as the common construct in exploring ways to apply the themes to the human-AI mode of writing. Both the extraction and inductive synthesis were conducted on a shared visual workspace, provided in Appendix~\ref{appendix:b}. The results of the theory/framework extraction are presented in Table~\ref{tab:papers1} and Table~\ref{tab:papers2}.

\begin{table*}[t]
\renewcommand{\arraystretch}{2} 
\tabcolsep=0pt
\begin{tabular*}{\textwidth}{@{\extracolsep{\fill}}p{0.28\textwidth}p{0.68\textwidth}@{}}
\textbf{Study} & \textbf{Theory/framework} \\
\hline
Buschek (2024)~\cite{Buschek2024-ka} & Proposes a new framework for AI writing by building on the construct of literary collage\\
Chakrabarty et al. (2024)~\cite{Chakrabarty2024-cu} & Applies the Cognitive Process Theory of Writing to design a collaborative writing tool\\
Cress and Kimmerle (2023)~\cite{Cress2023-rq} & Applies group cognition theory to explore writing and knowledge construction with AI\\
De Silva (2007)~\cite{De-Silva2007-jx} & Applies narrative-based writing and Rhetorical Structure Theory to prototype a collaborative writing tool\\
Dhillon et al. (2024)~\cite{Dhillon2024-qi} & Applies scaffolding theory to design an AI writing tool\\
Duin (1990)~\cite{Duin1990-uk} & Proposes terminology for collaborative writing, particularly for ways of working as a team\\
Ede and Lunsford (2006)~\cite{Ede2006-wh} & Proposes the hierarchical and dialogic modes of collaborative writing\\
Ens et al. (2011)~\cite{EnsUnknown-kv} & Builds on social constructivist theory and Ede and Lunsford's dialogic/hierarchical mode of writing to analyze writing processes\\
Fiore and Wiltshire (2016)~\cite{Fiore2016-vq} & Builds on team cognition theories to examine the role of technology in external cognition\\
Flower and Hayes (1981)~\cite{Flower1981-ey} & Proposes the Cognitive Process Theory of Writing\\
Gero et al. (2023)~\cite{Gero2023-nt} & Proposes a taxonomy of support dynamics that writers seek\\
Hayes (1996)~\cite{Hayes1996-ho} & Extends the Cognitive Process Theory of Writing to include social and physical aspects\\
Laban et al. (2024)~\cite{Laban2024-yv} & Proposes a three-stage information verification framework (Warn--Verify--Audit) to assist revision of AI-generated text\\
Larsen-Ledet et al. (2020)~\cite{Larsen-Ledet2020-nz} & Builds on Lowry et al.'s taxonomies of collaborative writing to add ``text functions'' as a way to incorporate the roles produced text takes in mediating collaboration\\
\hline
\end{tabular*}
\vspace{10pt}
\caption{Overview of included studies and the theories/frameworks used (Part 1; continued below). The table is sorted alphabetically according to the lead author's surname.}
\label{tab:papers1}
\end{table*}

\begin{table*}[t]
\renewcommand{\arraystretch}{2} 
\tabcolsep=0pt
\begin{tabular*}{\textwidth}{@{\extracolsep{\fill}}p{0.28\textwidth}p{0.68\textwidth}@{}}
\textbf{Study} & \textbf{Theory/framework} \\
\hline
Lay and Karis (1994)~\cite{LayUnknown-jh} & Discusses collaborative writing in industry through the lens of Bruffee and Bakhtin's concept of knowledge as an active, externally determined sensemaking process\\
Lee et al. (2024)~\cite{Lee2024-wa} & Proposes a structured taxonomy for AI collaborative writing tools\\
Li et al. (2024)~\cite{Li2024-ap} & Applies a user-centred design framework for collaborative writing and proposes ``reader-centred writing''\\
Lowry et al. (2004)~\cite{Lowry2004-me} & Proposes a foundational taxonomy for collaborative writing\\
Lowry and Nunamaker (2003)~\cite{Lowry2003-oo} & Proposes a taxonomy for online collaborative writing strategies\\
Mendoza-Chapa et al. (2000)~\cite{Mendoza-Chapa2002-lp} & Synthesises various group-awareness frameworks for collaborative writing tools\\
Posner and Baecker (1992)~\cite{Posner1992-ff} & Proposes a taxonomy of collaborative writing processes\\
Prasetyo and Bandung (2015)~\cite{Prasetyo2015-ne} & Applies Lowry et al.'s taxonomies of collaborative writing to build a collaborative writing tool\\
Rahhal et al. (2008)~\cite{Rahhal2007-mr} & Applies Rhetorical Structure Theory to support semantic consistency checking\\
Ruan et al. (2024)~\cite{Ruan2024-ri} & Proposes the Re3 framework for joint collaborative document revision\\
Wan et al. (2024)~\cite{Wan2024-rd} & Builds on the Cognitive Process Theory of Writing and proposes CoCo Matrix to recognize human effort in collaborative writing\\
Wiethof et al. (2021)~\cite{Wiethof2021-un} & Applies social response theory and the uncanny valley to build an AI collaborative writing tool\\
Wilbers et al. (2024)~\cite{Wilbers2024-kl} & Proposes the Overall Writing Effectiveness framework to assess human--LLM writing\\
Zhang et al. (2023)~\cite{Zhang2023-ku} & Applies Toulmin's argument model and the Cognitive Process Theory of Writing to build an AI writing tool with real-time logic visualizations\\
\hline
\end{tabular*}
\vspace{10pt}
\caption{Overview of included studies and the theories/frameworks used (Part 2; see Part 1 above).}
\label{tab:papers2}
\end{table*}

To reduce selection bias, two reviewers (the first and second author) took the same 15 randomly selected entries and applied the inclusion/exclusion criteria screening to check agreement levels. This sample screening not only revealed substantial agreement between the 2 reviewers (Cohen's $\kappa = 0.81$), it also helped to add specificity to the exclusion criteria, which the first author applied to the rest of the entries. The first author conducted the full-text reading and the data extraction alone, with the second and third authors helping to resolve any conflicts and achieve consensus.

\section{Results}

We now introduce the seven insights we drew from our review, along with design implications for future human-AI collaborative writing tools. Each design implication are supported with speculative interface sketches intended to illustrate how these implications can be instantiated, but these sketches have not been tested yet. The results are presented in Table~\ref{tab:results}.

\begin{table*}[t]
    \centering
    \renewcommand{\arraystretch}{2} 
    \begin{tabular*}{\textwidth}{@{\extracolsep{\fill}}p{0.5\textwidth}p{0.5\textwidth}@{}}
        \textbf{Themes from human-human collaborative writing} & \textbf{Design implications for human-AI collaborative writing} \\
        \hline
        Writing processes are iterative, non-linear and interchangeable any time
        &
        Use ``prototyping'' approaches (small, inspectable text chunks) to maintain cognitive alignment during delegation
        \\
        Supporting planning has been the primary focus in collaborative writing tools
        &
        Human writers are in need of more scaffolding around revision
        \\
        Group awareness is a key to success
        &
        Provide a moment for the human writer to articulate their own intentions before delegating
        \\
        Coherence support is an ongoing challenge
        &
        Visual representations of the document's semantic structure can scaffold human-AI alignment during iterative delegation
        \\
        Writing is socially constructed by readers
        &
        Generate and review documents with reader profile awareness
        \\
        Lack of clarity with authorship
        &
        Design data structures and visual cues to reinforce accountability
        \\
        Writing tools serve as external cognition
        &
        AI tools can contribute to not just task (writing) but process (coordination, sense-making, conflict resolution)
        \\
        \hline
    \end{tabular*}
    \vspace{10pt}
    \caption{Overview of the seven insights we drew from our review, along with design implications for future human-AI collaborative writing tools.}
    \label{tab:results}
\end{table*}

\subsection{Writing processes are iterative, non-linear and interchangeable any time}
\label{insight-nonlinear}

Cognitively, writing processes are highly flexible and non-linear. This is one of the key insights of the Cognitive Process Theory of Writing~\cite{Flower1981-ey}, which was the first to model the cognitive process of writing and has been referred to by many collaborative writing studies since. They proposed a hierarchically-organized model that primarily consists of 3 phases: (1) planning; (2) translating (i.e. writing); and (3) reviewing, with the monitoring process functioning as an orchestrator that decides when the writer moves from one process to another (Figure~\ref{fig:teaser}). A key aspect of this theory is that writers ``do not march through these processes in a simple 1, 2, 3 order''~\cite{Flower1981-ey} - rather, it is a constant iteration of creating goals, writing towards that goal, reading what was written, revising goals, writing towards the new goal, and so forth. Therefore, key subprocesses such as idea generation, revision and evaluation are interruptible by other processes at any given moment during the writing process. More simply put: writers revise as they write, plan as they revise, and write as they plan. Contrary to many beliefs, revision is not something that happens after writers do their writing, but rather something that happens ``while in the midst of generating a text''~\cite{Faigley1981-bt}.

This non-linear and interchangeable nature of the writing process was a common theme across various theories and frameworks. When explaining their proposed model of hierarchical and dialogic modes of writing, Ede and Lunsford~\cite{Ede2006-wh} equally stress the importance of resisting the urge to treat writing as a simple dichotomy between the two: ``it is inherently mixed and paradoxical''. Even prior to synchronous writing tools such as Google Docs, collaborative writing tools have supported parallel-partitioned writing, allowing members to view and edit each other's work at any given moment, which is known to increase teamwork effectiveness~\cite{Lowry2003-oo}.

\paragraph{\textbf{Design implication \#1: Use ``prototyping'' approaches (small, inspectable text chunks) to maintain cognitive alignment during delegation}}

While our review included studies that build on the three stages of the Cognitive Process Theory of Writing with AI~\cite{Chakrabarty2024-cu, Zhang2023-ku}, we argue that such literal implementations act as a hindrance to the iterative interchangeable nature of the human cognitive process. As explained in Section~\ref{distributed-cognition}, it is important to frame human-AI writing as a distributed cognitive system, but where the distribution of cognitive labor is currently unbalanced. Based on the Flower/Hayes model, we assert that the delegation of composition results in the human writers being pushed out of the cognitive loop, resulting in reduced ownership and error monitoring~\cite{Kosmyna2025-mp}.

\begin{figure*}[!t]
  \centering
  \includegraphics[scale=0.25]{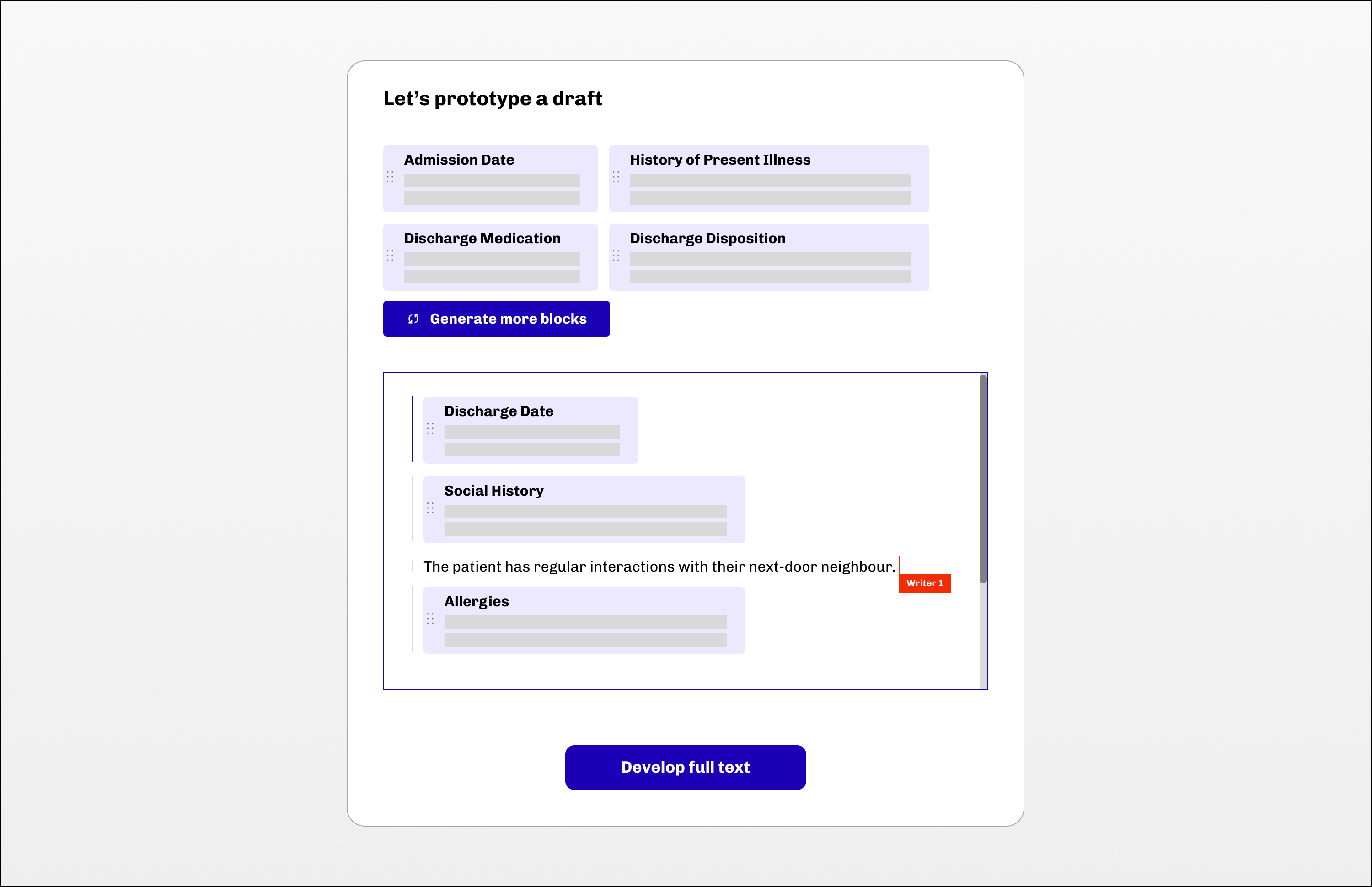}
  \caption{Example interface for design implication \#1: Use ``prototyping'' approaches (small, inspectable text chunks) to maintain cognitive alignment during delegation.}
  \Description{An interface showing small chunks of AI generated text, which the user can drag and drop into a separate field as well as add their own blocks and prototype the text before hitting the "develop full text" button.}
  \label{fig:implication1}
\end{figure*}

Rather, AI tools could be designed to support the optimal distribution of cognitive labor so that humans remain cognitively coupled even amidst AI delegation. This is where the analogy of prototyping could come in. In a prototyping process, the speed at which you iterate is key: you build something rough, test quickly, learn what could be improved, and move on to the next cycle. We argue that AI writing tools could encourage tighter cognitive alignment if they support the ``prototyping'' of text. For instance, imagine an interface like Figure~\ref{fig:implication1} where the text generation is intentionally provided as bite-sized, inspectable chunks which the human writer can read and review with little time and effort, compared to a fully developed paragraph. The human writer can directly reorder those chunks, add additional text or instructions for further refinements and run the generation again to retrieve another round of digestible chunks, thereby ''prototyping'' the text until they are satisfied, then hit the ``Develop Full Text'' button for the final version. Buschek's work on comparing the current human-AI collaborative writing scene with literary collage from avant-garde literature~\cite{Buschek2024-ka} frames such approach eloquently: writing with LLMs is very much the assembling of ``prefabricated text''.

Such prototyping approaches can potentially mitigate the loss of critical thinking~\cite{Lee2025-qu} of human writers that already see in essay writing tasks~\cite{Kosmyna2025-mp} or with image generation tools~\cite{Wadinambiarachchi2024-yy}. If we are to blindly generate documents or emails to send to our peers hardly reviewing their output, the results would be catastrophic, both for our cognitive process, and for workplace collegiality and responsibility. We argue that the problem here is not the ease of text generation, but the difficulty in reviewing and revising fully developed text. We see this to be a highly interesting and novel challenge that we welcome researchers to build further.

\subsection{Supporting planning has been the primary focus in collaborative writing tools}
\label{insight-planning}

Traditionally, planning has been the primary focus in CW tools~\cite{De-Silva2007-jx, Zhang2023-ku, Lowry2003-oo}. Prasetyo and Bandung discuss the planning features of their e-book writing tool in depth, but features for reviewing hardly go beyond simple annotations and comments~\cite{Prasetyo2015-ne}. This is understandable, since revision was primarily a human-centric task: it requires humans to read the work that they wrote. However, in cases where the LLMs take the role of writer, we believe that some emphasis should shift from planning as a function of the tool, to support other activities. Indeed, in Chakrabarty et al.'s study on supporting professional writers with LLMs~\cite{Chakrabarty2024-cu}, the AI contributions with the lowest retention rate (i.e. not kept until the final document) were those related to high-level planning such as ideas for a coherent plot progression, while the highest retention (i.e. kept in the final document) were those related to translation such as a rewrite of a particular passage. Perhaps this is reflecting one of the core differences between human and AI writers: AI tools do not have a stake in the written content, and therefore it is seemingly unfruitful to plan together. The strength of AI lies not in the rigorous commitment to the message conveyed, but rather in the faster and easier translation of ideas to words, and the meticulous revision of errors. Interestingly, back when we made the shift from writing on paper to writing with word processors, studies showed that undergraduate writers planned less with word processors than with pen and paper~\cite{Hayes1996-ho, Haas2013-av}. If the shift from pen and paper to word processors reduced the (presumed) need for planning, then it seems plausible to assume that there will be an even larger reduction when shifting from manual writing to LLMs. That is not to say, however, that the value of planning is diminishing. Planning stays just as important, but from the standpoint of appropriate distribution of cognitive labor, it makes more sense for the AI to prioritize reviews over planning.

\paragraph{\textbf{Design implication \#2: Human writers are in need of more scaffolding around revision}}

Now that the generation of text itself can be done by LLMs effortlessly within seconds, we may encourage users to get something down on paper first, then revise and plan accordingly, nudging writing more towards an empirical ``trial and error'' process than a rigorously planned out one, again similar to prototyping. This suggests opportunities for writing tools to shift its focus from planning to revising (of AI generated text).

Though outside of this review, Gordon et al.~\cite{Gordon2023-li} calls such approach "co-audit", where the AI contributes to not just the text generation, but also the human oversight in a human-in-the-loop model. They propose the ``prompt-response-audit'' cycle which share a similar ethos as our ``plan/delegate/review'' model. From our review, Laban et al.~\cite{Laban2024-yv} proposed the Warn-Verify-Audit framework, where new information introduced to the document by LLMs are constantly flagged, and are automatically searched online or sent to an external auditor for manual fact checking. We see similar interesting fact-checking approaches emerging~\cite{Liu2024-ht, Amirizaniani2024-dx, Dietz2024-jz}, all of which are AI-enhanced human-in-the-loop approaches to maximize the strengths of manual oversight and LLM automation.

\begin{figure*}[!t]
  \centering
  \includegraphics[scale=0.65]{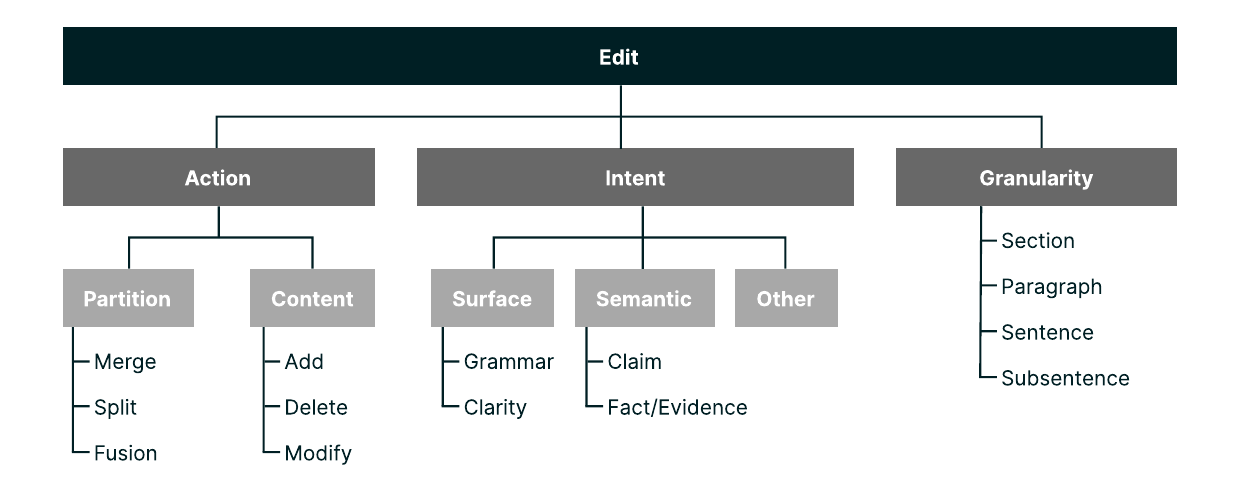}
  \caption{Taxonomy of revision proposed by Ruan et al.~\cite{Ruan2024-ri}}
  \Description{A diagram of granular types of revision, as proposed by Ruan et al.}
  \label{fig:re3}
\end{figure*}

\begin{figure*}[!t]
  \centering
  \includegraphics[scale=0.25]{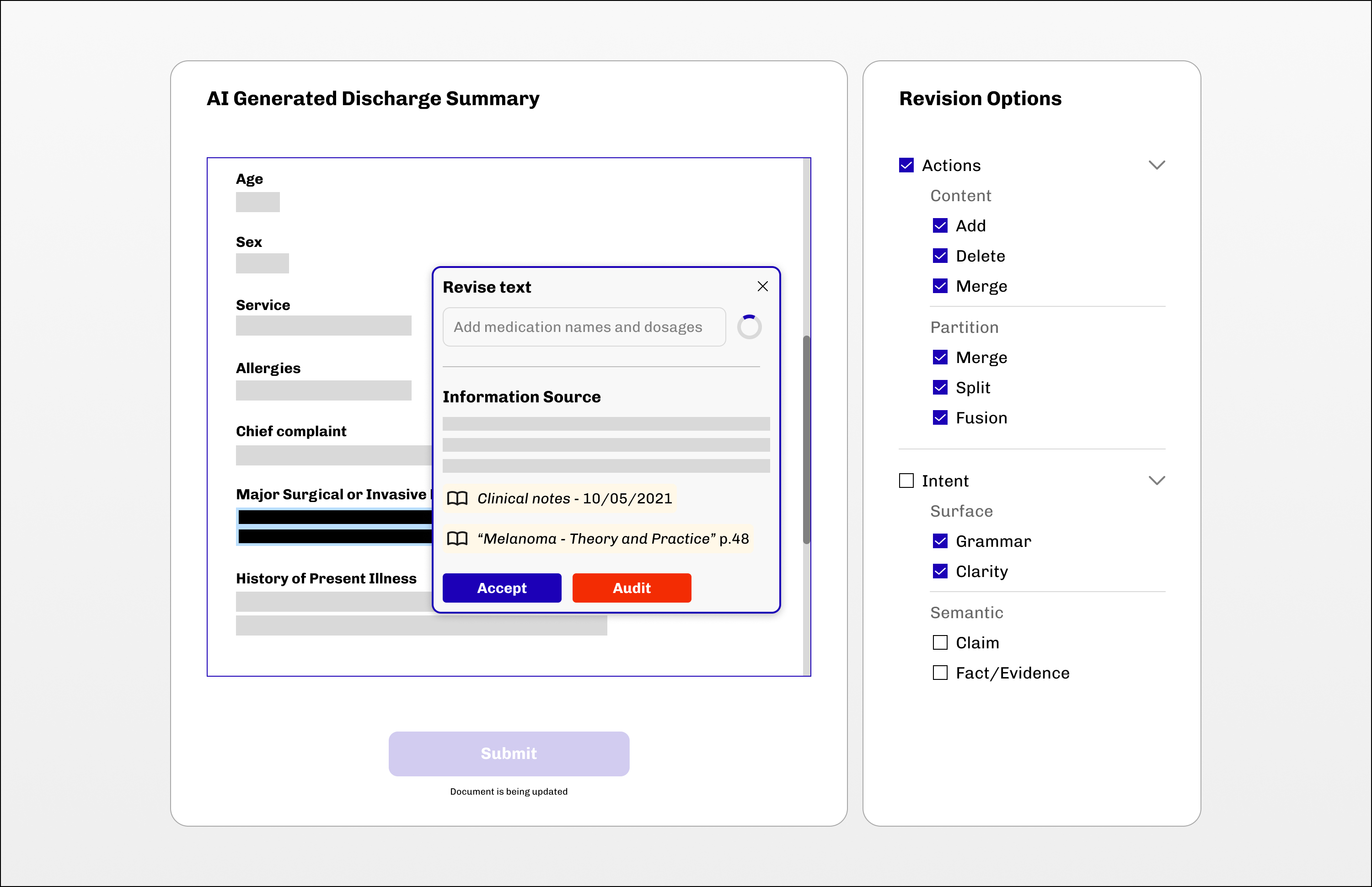}
  \caption{Example interface for design implication \#2: Human writers are in need of more scaffolding around revision.}
  \Description{An interface showing various AI support for text revision, such as specifying the granular types of revision, built in fact checkers and partial regeneration of highlighted text.}
  \label{fig:implication2}
\end{figure*}

Shifting the focus from planning to revision, however, also comes with a cost. LLMs are prone to sycophancy bias~\cite{Sharma2023-bv}, essentially making it a ``yes man'' regardless of factual truth. If so, is it unrealistic to expect genuine critical feedback from LLMs? Hints can be drawn from the Re3 framework~\cite{Ruan2024-ri}, which proposes the taxonomy of revision as shown in Figure~\ref{fig:re3}. By dividing revision by its qualitative dimensions such as action, intent and granularity, and further subdividing each by their characteristics, we could enable users to provide more granular control over revision instructions, such as requesting for grammar or clarity related edits for intent on sentence or subsentence granularity, but preventing factual content updates which LLMs are known to be unreliable with. Interfaces like Figure~\ref{fig:implication2} could contribute to the redistribution of cognitive labor to emphasize reviewing, and further feed the revisions into the next cycle of planning. Given that the revision has long been mistakenly seen as just the ``tidying-up activity'' after completing a first draft~\cite{Faigley1981-bt}, helping the human writers to get accustomed to these more granular terminologies around revision is perhaps the key agenda here.

\subsection{Group awareness is a key to success}
\label{insight-groupawareness}

Because human-human collaborative writing relies heavily on the ability to reach consensus, group awareness has been seen as a key to its success~\cite{Mendoza-Chapa2002-lp}. Group awareness in collaborative writing is not just about having awareness of what other members are doing during the collaboration process. Lowry et al.~\cite{Lowry2004-me} break down the types of group awareness for collaborative writing into informal awareness (knowing \textit{where} the other members are working), group-structural awareness (knowing the roles and responsibilities of other members), social awareness (knowing the motivation and emotional state of other members) and workspace awareness (knowing what other members are up to in the software in use). Similarly, Duin~\cite{Duin1990-uk} state the importance of sharing ``terms'' when writing collaboratively. This includes highly human traits such as sharing each writer's preferred writing process, commitment levels and skills prior to commencing the collaboration process, deciding on when to stop revising, and ending the collaboration session with appreciation of each other's work.

\paragraph{\textbf{Design implication \#3: Provide a moment for the human writer to articulate their own intentions before delegating}}

As stated in Section~\ref{scope}, we prefer to see AI simply as a tool and not a team mate~\cite{Naikar2025-sq}. While the use of human-like metaphors are certainly of benefit in human-AI collaboration, anthropomorphizing AI as a team mate often misleads users to false expectations, which is a problem when AI simply cannot be accountable for written content~\cite{Stokel-Walker2023-ry}. Therefore we consider straightforward translations of human-human group awareness to human-AI collaboration (such as considering the AI's motivations, emotional state or preferred writing styles) to be excessive anthropomorphism. However, the field of Explainable AI has also shown that uncovering the human's expectation for AI tool's roles and boundaries prior to its use could help mitigate under- and over-reliance. For instance, Cai et al~\cite{Cai2019-yj} found that, in clinical contexts, clinicians wished to know the training data, the medical views and the design objectives of the AI, far before engaging in the explanations of their diagnosis recommendations. As long as it is in service of improving the user's motivation towards the task at hand, it may be worthwhile to provide an onboarding moment to share the user's expectations on the AI tool's roles and capabilities, prior to commencing the writing. We see Wiethof et al.~\cite{Wiethof2021-un}'s work to be in line with this principle: they provided users with a written self-introduction by the AI agent prior to the writing, which led to ``realistic expectations''. However, their implementation also came with a rather literal adoption of Nass and Moon's Social Response Theory~\cite{Nass2000-qi} by providing social identities to the agent, which we see to be an example of excessive anthropomorphism. Similarly, Cress and Kimmerle~\cite{Cress2023-rq} suggest refining human-human group awareness to suit the more asymmetrical human-AI relationship, and propose AIs to develop \textit{theory of mind}, the cognitive ability to be aware of others' mental states. A good balance here may be to simply provide a moment for the writer to individually reflect on what support they are seeking from the AI co-writer, to nurture their own self-awareness, but stay away from introducing the AI co-writer as another team mate with its own values and writing styles.

\begin{figure*}[!t]
  \centering
  \includegraphics[scale=0.25]{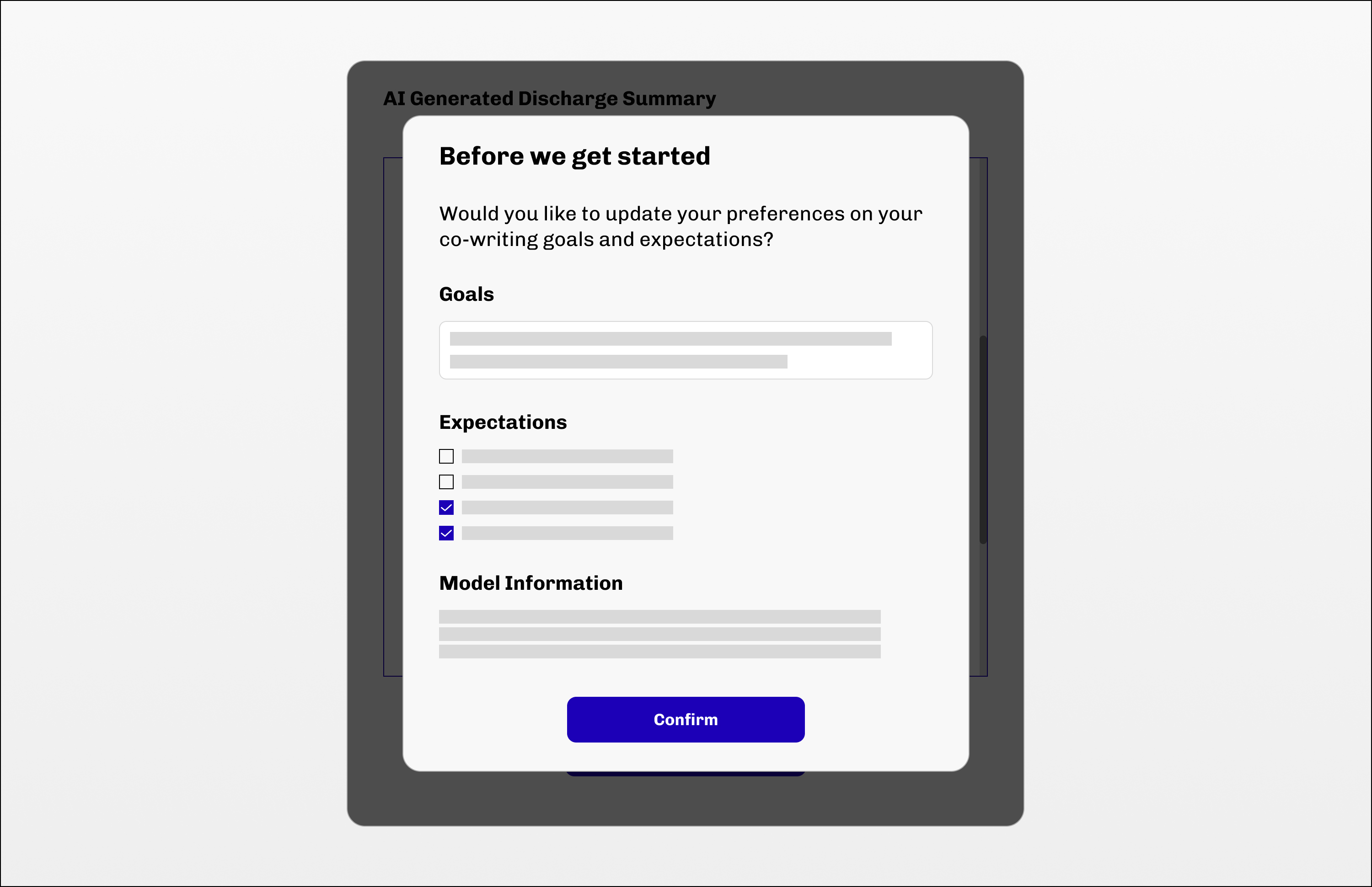}
  \caption{Example interface for design implication \#3: Provide a moment for the writer to articulate their own intentions before delegating.}
  \Description{An interface showing the onboarding screen before collaborative writing with AI, asking users to articulate their goals and expectations from the AI.}
  \label{fig:implication3}
\end{figure*}

From a distributed cognition lens, this distinction can be framed through the separation of socially distributed cognition and external cognition. That is, using an AI tool to reflect and articulate one's writing goals and motivations is a suitable and effective example of external cognition, especially given the fact that LLM interactions are much more difficult than traditional software in that users are required to articulate their ``intentions''~\cite{Subramonyam2024-sd}. However, expecting the AI to behave as a team member would be an unsuitable example of socially distributed cognition, since current AI tools are neither accountable nor have their on intentions for the writing task. As such, the cognitive scaffolding that the AI tool provides to support group awareness must be designed carefully to only support external cognition, and not mistakenly lead the user in thinking that a socially distributed cognition can be achieved with it, such as the example shown in Figure~\ref{fig:implication3}.

\subsection{Coherence support is an ongoing challenge}
\label{insight-coherence}

When writing with others, it is crucial to have syntactic and semantic consistency between the different parts of the documents. This has been an ongoing challenge in collaborative writing, particularly in the CSCW context~\cite{De-Silva2007-jx}. Given that consensus and group awareness are key factors in human-human collaborative writing, it is understandable that coherence becomes a vulnerable factor that tends to get lost in the complexity of human collaboration, particularly with creative writing~\cite{Wan2024-rd}. Studies that tackled this in our review~\cite{De-Silva2007-jx, Rahhal2007-mr} both used the Rhetorical Structure Theory to semi-programatically point out semantic incoherence in documents. In Rhetorical Structure Theory, the document is first broken into non-overlapping segments, and the relationships between each segment are categorized into pre-defined labels. Some relationships resemble equal importance, while some resemble a hierarchical relation. Given that such semantic classification is one of the key strengths of LLMs, we believe that, if designed appropriately, AI co-writer tools could support collaborative writing without jeopardizing coherence much better than human team mates.

\paragraph{\textbf{Design implication \#4: Visual representations of the document's semantic structure can scaffold human-AI alignment during iterative delegation}}

\begin{figure*}[!t]
  \centering
  \includegraphics[scale=0.25]{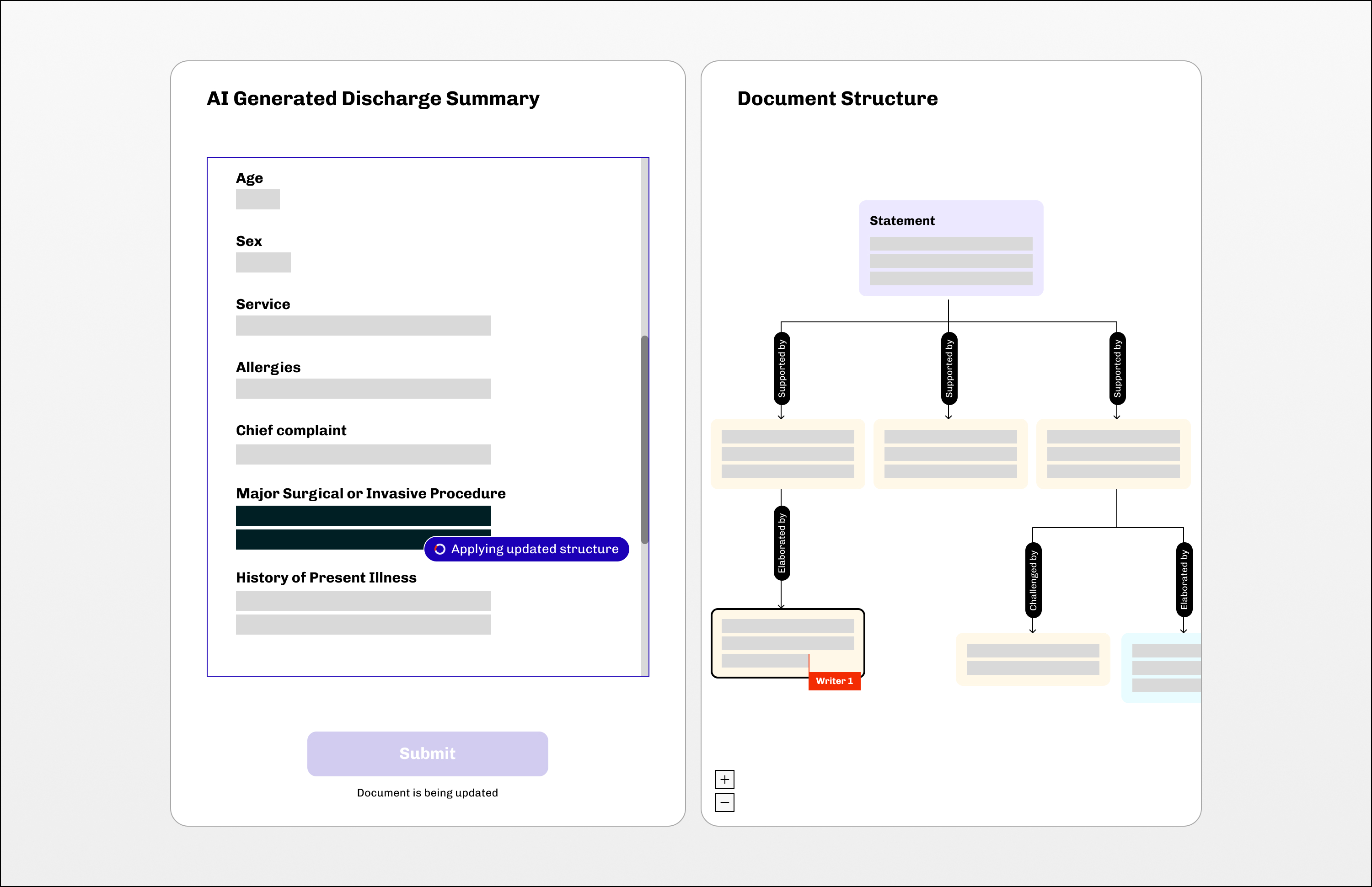}
  \caption{Example interface for design implication \#4: Visual representations of the document's semantic structure can scaffold human-AI alignment during iterative delegation.}
  \Description{An interface showing a real-time visualization of the semantic structure of the written text, using nodes and connectors.}
  \label{fig:implication4}
\end{figure*}

While appropriate prompting could potentially be a solution for achieving lexical (i.e. word level) and syntactic (i.e. grammar level) alignment, semantic alignment (i.e. consistency in the flow and document structure) would require a more intricate design intervention, since it is less about the LLM's writing capabilities and more about human-AI alignment. From a distributed cognition lens, this is a great example of externalised cognition, and is something that has been technologically difficult prior to LLMs. Visual representation of the semantic document structure as seen in VISAR~\cite{Zhang2023-ku} serves as a great inspiration here. Interfaces like Figure~\ref{fig:implication4} that visualizes the current document structure in real-time could enable the user to have an understanding of ``how their text {\emph{works}}''~\cite{Rahhal2007-mr}, and scaffold constant alignment with the AI. Further, it could also contribute to the scaffolding of review discussed in Section~\ref{insight-planning}, by making it easier to grasp the changes made to the arguments throughout the iterative development.

\subsection{Writing is socially constructed by readers}
\label{insight-readers}

The notion that writing is socially constructed by the reader has been a key aspect of studies on writing and rhetoric in general. Bahktin has been a key figure here, stating that writing is a form of dialogue, and is a passive act that is socially interpreted by the readers in an ``ever changing context''~\cite{LayUnknown-jh}. Interestingly, when Hayes updated the original Cognitive Process Theory of Writing, they added the \textbf{reader} as a social environment of writing~\cite{Hayes1996-ho} as well. Some studies in our review took a rather literal approach to this socially-constructed aspect of writing. For example, Li et al.~\cite{Li2024-ap} suggests incorporating readers into the writing process by applying the concept of ``user-centered design'' to writing, proposing a ``reader-centered writing'' process where the text is automatically adapted to the readers' profiles.

\paragraph{\textbf{Design implication \#5: Generate and review documents with reader profile awareness}}

\begin{figure*}[!t]
  \centering
  \includegraphics[scale=0.25]{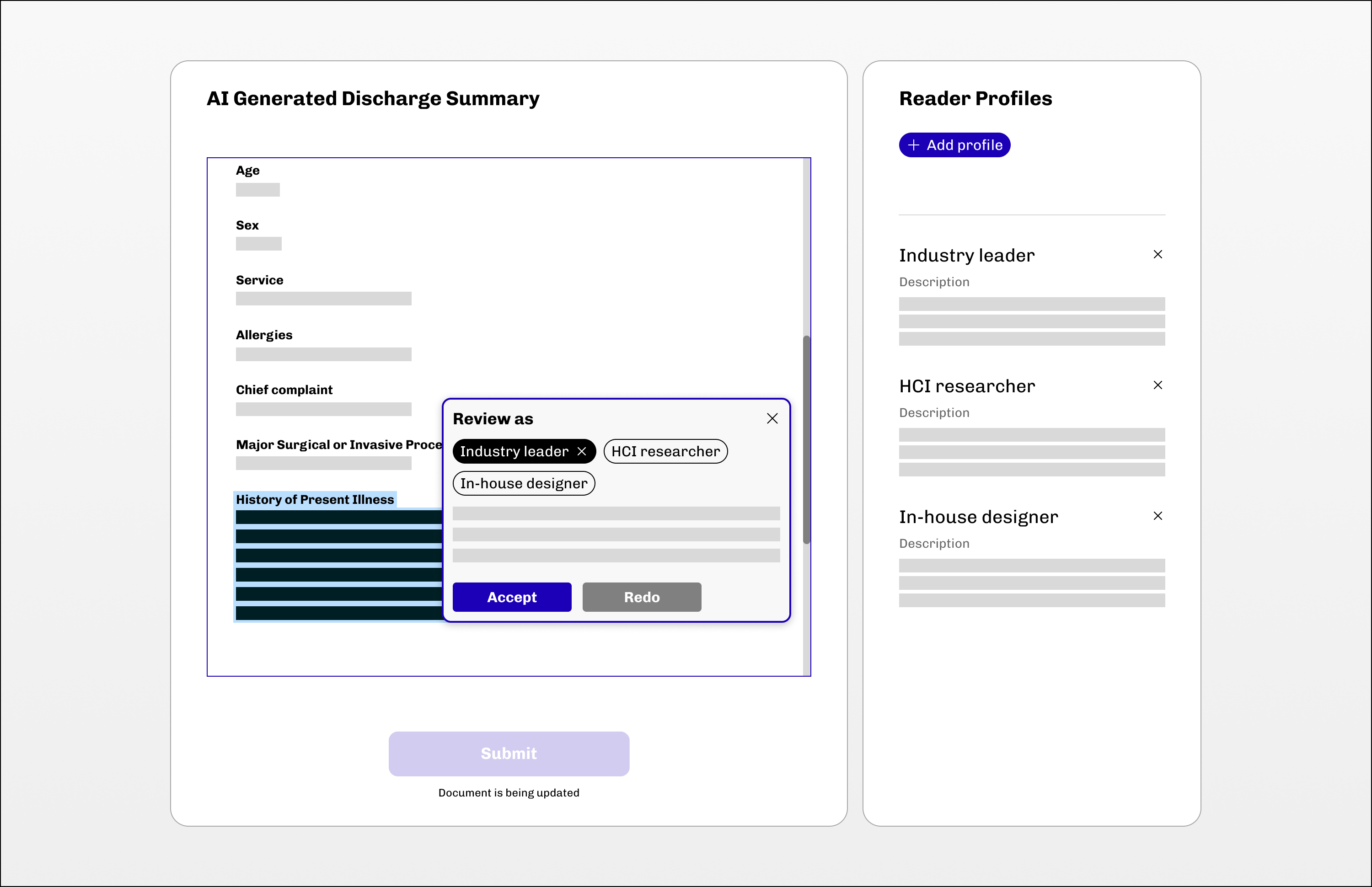}
  \caption{Example interface for design implication \#5: Generate and review documents with reader profile awareness.}
  \Description{An interface showing various potential reader profiles on the right panel, which can be selected to review the written text based on that profile.}
  \label{fig:implication5}
\end{figure*}

If readers are a part of the writing, then it would mean that readers are inseparable from the LLM's text generation process. A straightforward implementation would be to simply include the intended audience profile to the prompt, but there is more that can be done. The fundamental implication here is that writers benefit from being conscious of the reader's grammar, language and existing knowledge to maximize the ease of comprehension~\cite{Rahhal2007-mr}. One possible approach is to incorporate reader profile awareness into the \textbf{revision} process, where the AI co-writer is given the ability to assess the text as a pre-defined audience to check for any inconsistencies or lack of information from their perspective, like the interface in Figure~\ref{fig:implication5}. The risk here would be excessive personalization, leading to bias amplification and social echo chambers, like we are starting to see in the adjacent field of human-LLM dialogue~\cite{Knoeferle2025-dh}.

\subsection{Lack of clarity with authorship}
\label{insight-authorship}

The concept of authorship has been a particularly debated topic in collaborative writing~\cite{Ede2001-zd}. While we excluded papers that primarily focused on authorship during our screening, it has still come up as an evident theme in the reviewed papers. For instance, Dhillon et al.~\cite{Dhillon2024-qi} explored the effects of scaffolding during human-AI collaborative writing, and found that high scaffolding (i.e. suggestions for entire paragraphs) resulted in higher writing quality and productivity, but also resulted in the decrease of text ownership. As more and more writing is conducted by AI tools, it seems evident that our sense of ownership towards the text will decrease~\cite{Kosmyna2025-mp}. Given that the concept of authorship is directly tied with the author's accountability~\cite{Stokel-Walker2023-ry}, and given that only the human writers can be kept accountable for the consequences of any human-AI collaborative writing~\cite{Stokel-Walker2023-ry}, it seems crucial for AI writing tools to contribute to the clarification of accountability.

\paragraph{\textbf{Design implication \#6: Design data structures and visual cues to reinforce accountability}}

\begin{figure*}[!t]
  \centering
  \includegraphics[scale=0.25]{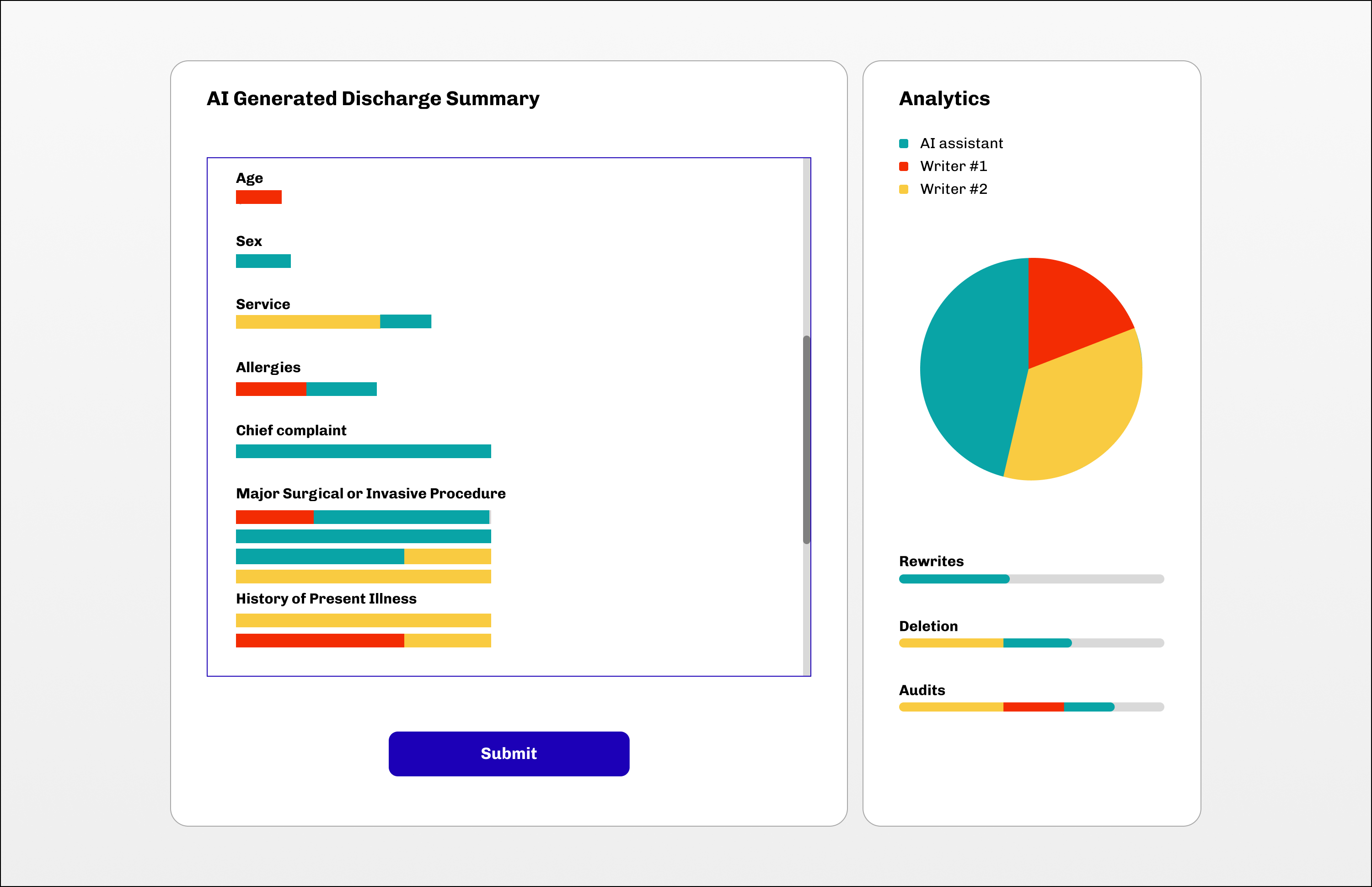}
  \caption{Example interface for design implication \#6: Design data structures and visual cues to reinforce accountability.}
  \Description{An interface that distinguishes the original writer of a text, with the text being color coded in various colors. The right panel is showing the analytics such as contribution ratio.}
  \label{fig:implication6}
\end{figure*}

Our review points towards the distinction of human contribution v.s. LLM contribution as an effective strategy here. Laban et al's~\cite{Laban2024-yv} InkSync proposes an insightful JSON data structure that can constantly keep track of all the text revisions that were applied to the document, and what interactions originated the edit (such as chat, comments, or their built-in brainstorming feature). Further, it checks whether the newly added text is introducing new information to the document, which can be automatically flagged for verifications via search engines or external audits. Such data structures would allow collaborative writing tools to continuously distinguish human written text from AI written ones, enhancing the notion of accountability for the user. In Kosmyna et al.~\cite{Kosmyna2025-mp}'s study on essay writing with ChatGPT, they also touch on how lower ownership of leads to the ``diminished sense of cognitive agency''. That is, when we feel that a written text is AI's and not ours, we will feel less control over the product, leading to decrease in error monitoring. This suggests opportunities for going beyond visualization of writer distribution, and promoting human writers' self awareness of their sense of ownership over the text, such as visualizing contribution ratios and edit types (Figure~\ref{fig:implication6}).

\subsection{Writing tools serve as external cognition}
\label{insight-externalcognition}

Fiore and Wiltshire~\cite{Fiore2016-vq} explored the various theories related to technology as an \textit{artifact of external cognition} to make the claim that tools (be it AI or not) are inseparable from a team's cognition. Of particular interest is their claim that, even within external cognition, it is important to distinguish \textbf{teamwork} (behaviors required for working together with others) and \textbf{taskwork} (functions required for getting the task done). Tools can support teamwork through the visualization of group awareness, team progress and facilitating communication, while tools can also support taskwork through the direct assistance of the task (e.g. writing or decision making). 

\paragraph{\textbf{Design implication \#7: AI tools can contribute to not just task (writing) but process (coordination, sense-making, conflict resolution)}}

\begin{figure*}[!t]
  \centering
  \includegraphics[scale=0.25]{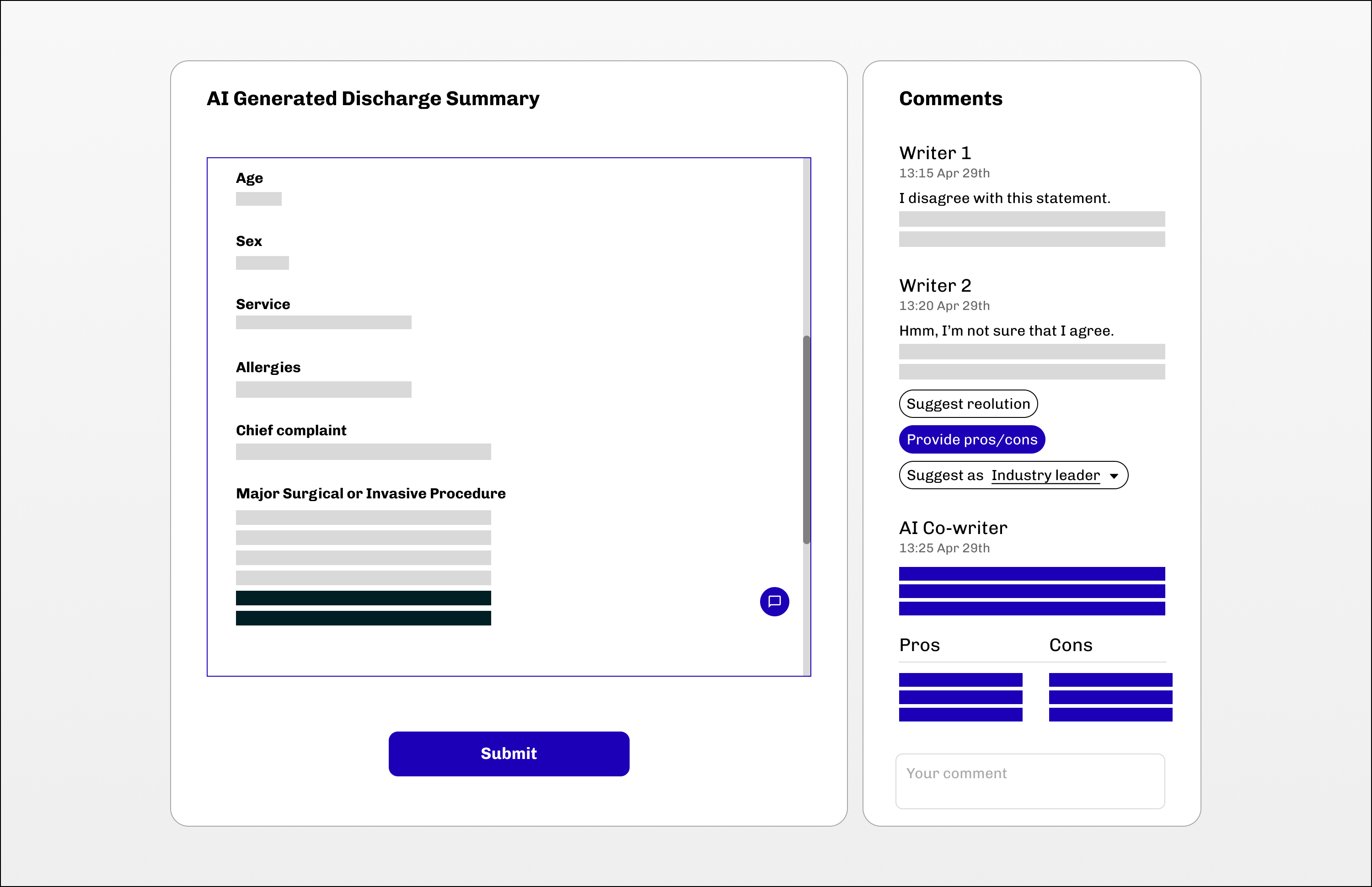}
  \caption{Example interface for design implication \#7: AI tools can contribute to not just task (writing) but process (coordination, sense-making, conflict resolution).}
  \Description{An interface showing an AI intervening with conflicting human opinions in the comments section, providing options for resolution methods.}
  \label{fig:implication7}
\end{figure*}

As we design AI writing tools, we must take great care in making sure that the tool is not resulting in additional cognitive load, leading to more disruption than improvement. The key learning here is that undesired cognitive load can be witnessed at various layers of the writing process, such as communicating with other stakeholders and checking in on their progress, not just in the composition itself. For example, Larsen-Ledet et al.~\cite{Larsen-Ledet2020-nz} shed light on writers' cognitive load from a socio-technical perspective, finding that writers often like to switch between multiple writing tools, not just for functional reasons, but to get away from the collaborative mode to ``sit for myself.'' That is, writers sometimes intentionally choose to stay away from collaboration to stay focused. These findings suggest that the AI's role in externalising cognition is not limited to the production of high quality text (taskwork), but also to make sure that the process itself is not causing discomfort for the user (teamwork). Interfaces like Figure~\ref{fig:implication7} could help to offload cognition from the teamwork perspective, by streamlining the collaboration process itself, both for human-AI writing as well as human-human teamwork.

\section{Discussion}

\begin{figure*}[!t]
  \centering
  \includegraphics[scale=0.7]{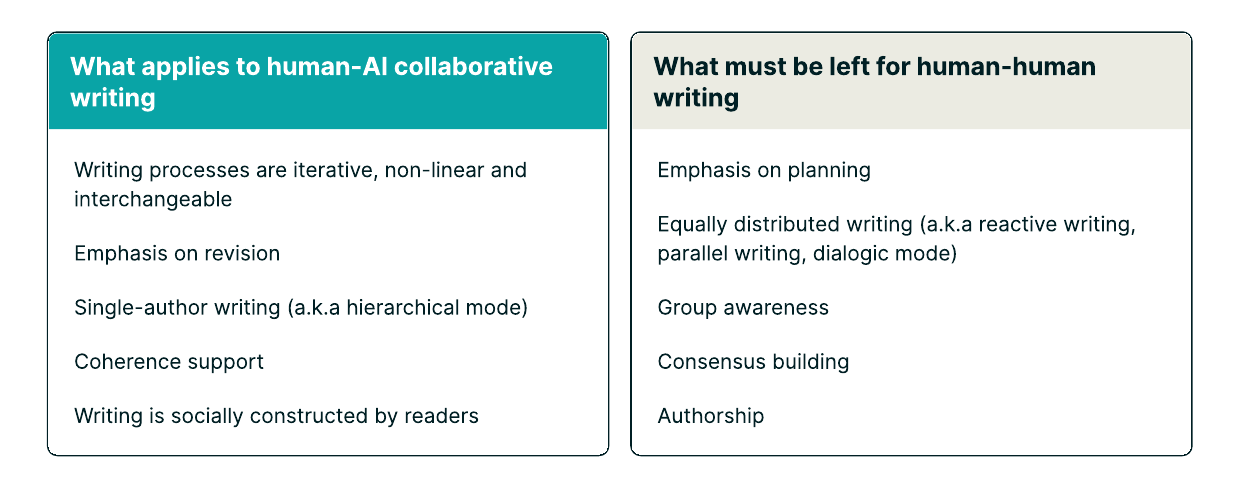}
  \caption{Theories/frameworks from human-human collaborative writing that apply to human-AI collaborative writing (left), and those that are less applicable (right).}
  \Description{A section on the left lists the theories and frameworks that still apply to human-AI collaborative writing, such as "emphasis on revision" and "single-author writing". The section on the right lists those that must be left for human-human writing, such as "emphasis on planning" or "equally distributed writing".}
  \label{fig:discussion}
\end{figure*}

We now come back to our initial research questions: what insights can we draw from theories/frameworks on human-human collaborative writing to apply to the design of human-AI writing tools? Which parts of those theories/frameworks are not relevant for human-AI collaboration? We provide an overview in Figure~\ref{fig:discussion}.

\subsection{Applying existing human-human collaborative writing theories/frameworks to the human-AI paradigm}

As more humans choose to delegate the actual writing to the essentially effortless AI generation, the role of humans shift from a hands-on writer to more of an editorial decision maker~\cite{Buschek2024-ka}, leading to a skewed redistribution of cognitive labor. However, given the non-linear, interchangeable nature of writing based on the Cognitive Process Theory of Writing~\cite{Flower1981-ey}, delegating the entire text composition process may push humans out of the cognitive loop, leading to reduced sense of ownership and error monitoring~\cite{Kosmyna2025-mp}. While cognitive offloading has clear benefits, we argue that the current unbalanced cognitive distribution must be redesigned to keep the humans \textbf{stay} in the loop when it comes to writing, to make sure that the human writer stays accountable for the writing they produce (which the AI cannot)~\cite{Stokel-Walker2023-ry}. For this we discuss scaffolding the human cognition through a ``prototyping'' approach to writing, where the AI enables rapid, small-scale iterations of text that keep writers mentally engaged even when the generation is delegated to AI. For instance, tools could enable quick iterations of the cognitive process by starting with inspectable, smaller chunks of text, and gradually moving on to a fully fleshed-out document, much like evolving prototypes from low-fidelity to high-fidelity. Consequently, further scaffolding will be required for the revision stage of the writing process, namely interactions to allow human writers to read, review and edit AI generated text quickly without overwhelming mental demand. For this, tools can equip writers with enhanced understanding and terminologies for various revision types, to set more granular boundaries for the desired AI support and to draw out the best of human and AI performance.

As AI takes on more of the content writing, the ongoing challenge of coherence support will be of further importance. Given that LLMs provide a level of semantic generalization and abstraction not achievable with prior AI models, we are optimistic about the impact that LLMs can bring to the alignment of document structures, such as through real-time visualization. These interventions could positively impact not just the human-AI collaborative writing but enhance human-human writing as well.

Since we take the stance that AI-writing assistants are a tool where humans remain in control, teamwork-related factors such as group awareness, consensus building and authorship, which have been much discussed in traditional collaborative writing studies, will be less applicable to the human-AI paradigm, except between multiple human co-authors. Therefore, when it comes to the writing strategies, we believe most human-AI collaborative writing will end up as single-author writing, or will be conducted in the hierarchical mode where the AI co-writer will primarily translate ideas to words instantaneously under human supervision, and humans will primarily take an editorial role. The more equally distributed forms of collaborative writing, known as reactive writing~\cite{Lowry2004-me}, parallel writing~\cite{Lowry2004-me} or the dialogic mode~\cite{Ede2006-wh}, will remain a strategy unique to human collaboration. However, given that the reduction of cognitive load can derive not just from the direct support of text production but from the streamlining of the overall writing process itself, it is equally important that the AI co-writer actively works to resolve frictions in the collaboration process. That is, the cognitive offloading can happen not just in the generation of text, but also in the \textbf{facilitation of the writing process}. We deem this notion to be of particular importance, since this could pave the way towards collaborative writing with AI without necessarily disrupting our sense of ownership or error monitoring.

Finally, the notion that writing is socially constructed through readers remains valid, and we urge researchers to explore the implications that LLM's diverse text generation capabilities will have to this space. At the same time, we acknowledge that this will only apply when humans are the readers: if human and AIs are co-writing, there is no reason to doubt that humans and AIs will co-read, and if documents are to be agentically processed in future workplaces, this will signal another significant turning point for writing in general.

Much like how word-processors had a profound impact on the way that we wrote with pen and paper, LLMs seem to be fundamentally shifting the way that we conduct writing. While there are certainly signs of de-skilling, as HCI researchers we are also curious to see what synergistic effects may arise from this shift, resulting in novel mindsets and methodologies of writing. As Brinkmann et al.~\cite{Brinkmann2023-lo} propose through their lens of ``machine cultures'', our culture of writing is already being reshaped by our interactions with LLMs.

\subsection{Prompting as a form of writing}
In Figure~\ref{fig:teaser}, we assumed that, when writing with AI, prompting becomes the key component of the planning process. When prompting LLMs, we consider not just the content of the writing but the audience, length, structure, tone, and internal knowledge that we posses (often referred to as ``context'' in prompt engineering) that is required for the writing. This is much in line with how Flower and Hayes describe their model: they use the term planning in a ``much broader sense'', where writers form abstract ``representations'' of ideas and goals, often involving retrieval from long-term memory~\cite{Flower1981-ey}. The way they explain the iterative nature of goal-setting is also highly analogous to the way LLMs are used today, refining instructions through multiple trial and errors~\cite{Arawjo2023-qi, Liu2022-an}.

We note here that, interestingly, \textit{prompting itself} is also a form of writing. That is, the act of writing a prompt is also the very result of the writer's cognitive writing process. In this scenario, the user would first engage in the \textit{planning/translating/reviewing} to write a prompt, which would then be fed to an LLM for the user to engage this time in a distributed \textit{plan/\textbf{delegate}/review} process. Further, if the writer chooses to use meta-prompting~\cite{Reynolds2021-nq} (using LLMs to create and refine prompts), then they go through the \textit{plan/write/review} process to write a prompt that retrieves a prompt which they would revise and iterate through the \textit{plan/\textbf{delegate}/review} process, then apply the generated prompt to the LLM to generate the final outcome which would be further revised through the \textit{plan/\textbf{delegate}/review} process, and so forth. In any case, however, when viewed from the lens of this study, prompt writing is distinctly different from most forms of writing in that it is a uniquely reader specific form of writing --in this case the reader being the LLM. Grammar or spelling errors are not rigorously checked, since they do not matter as long as the instruction gets through. Additional contexts are also often added without much thought, regardless of knowing whether the outcome would be improved from doing so \cite{Liu2022-an}. Prompt engineering or defense techniques~\cite{Liu2023-cs} (such as the ``sandwich prevention'', which refers to repeating the certain instructions at the beginning and the end of a prompt) are often highly unnatural forms of writing. Despite these peculiarities as a form of writing, it is also interesting that the concept of authorship seems to apply to prompts: platforms such as PromptBase\footnote{See \url{https://promptbase.com/}} allow users to sell and buy effective prompts. As such, prompting itself can be seen as the act of writing, with distinct characteristics. In this sense, when humans co-write with AI, the process begins with a legacy human mode of writing, resulting in a mix of both approaches that we outlined in Figure~\ref{fig:discussion}. How does this affect the overall process of writing with AI? Is it necessary to intentionally separate the initial prompting (i.e. the legacy \textit{plan/write/review} process) from the human-AI mode of writing (i.e. the \textit{plan/\textbf{delegate}/review} process) that follows? These are still early questions in trying to understand how LLMs are affecting the way that we write, and we welcome more researchers to explore this space.

\section{Limitations}
We acknowledge some general limitations of our literature analysis approach.

First, we chose to look at Web of Science and the ACM Digital Library, but it is possible that we are missing collaborative writing theories or frameworks that were not covered by these two databases. For example, among the 6 articles that we added through backward referencing, 4 were found to be in the ProQuest database (the remaining 2 were the cognitive process theories by Flower and Hayes~\cite{Flower1981-ey, Hayes1996-ho}, which are fundamental to this study but are technically not collaborative writing). We are, however, confident that we have sufficiently covered at least the most influential ones commonly found in past literature.

Second, our findings apply only to the scope of collaborative writing that the authors have chosen to study, as explained in Section~\ref{scope}. Our results will not be applicable to collaborative writing in other contexts, such as fictional writing or educational settings -- these may be explored in future work.

Third, we intentionally cast a wide net for the systematic review, trying to capture as many related theories and frameworks from as many disciplines as possible. This has led to the decision of not screening based on their impact or citations, which may have led to variations in the quality of the surveyed papers.

\section{Conclusion}
In this paper, we have argued for the need to reassess foundational theories and frameworks of collaborative writing to adapt to the human-AI paradigm, as LLMs increasingly become our daily co-writers. We conducted a critical review of existing theories and frameworks on human-human collaborative writing to assess their relevance to the human-AI paradigm, and drew seven insights and implications that can be applied to the design of human-AI writing tools. We found that, as we delegate more writing to AI, our cognitive process shifts from the traditional \textit{plan/write/review} process to a \textit{plan/\textbf{delegate}/review} process, pushing the humans out of the cognitive loop due to excessive offloading. To ensure that humans can stay cognitively aligned, we discuss scaffolding the human cognition through a ``prototyping'' approach to writing, where the AI enables rapid, small-scale iterations of text that keep writers mentally in the loop even when generation is delegated to AI. AI tools would also need to provide increased scaffolding for revision methodologies as well. In turn, this may reduce the focus on the planning aspects of collaborative writing due to the ``trial-and-error'' approach to writing, increasing the importance of coherence support between human and AI writing. We believe that the LLM's strong semantic capabilities could provide analyses that were previously difficult with traditional methods, such as real-time visualizations of a document's semantic structure. Human teamwork constructs such as group awareness, consensus building and shared authorship, which have been central discussions in human-human collaborative writing, would not apply to the human-AI paradigm, resulting in the majority of human-AI collaborative writing to be categorized as single-author writing strategy. Finally, writing will continue to be socially constructed by readers, but this is dependent on the fact that humans will continue to be the primary readers of the documents we create -- which may not necessarily be the case in the near future. Beyond theoretical insights, we provide practical design implications and speculative interface sketches to guide real-world application. We hope that our work will bring clarity and theoretical grounding to the interaction designs of human-AI collaborative writing.

\section{Conflicts of interest}
The authors declare that they have no competing interests.

\section{Data availability}
The data underlying this study are available within the article. The final extracted study list is included in the Results section, and the synthesis process is documented through exported Miro boards attached as supplementary material. The Excel file used for the PRISMA screening process is available from the corresponding author upon request.

\begin{acks}
The interface designs in this paper used Google's Material Icons\footnote{See \url{https://fonts.google.com/icons}} provided under the Apache License, Version 2.0.
\end{acks}

\bibliographystyle{ACM-Reference-Format}
\bibliography{sample-base}


\appendix

\section{Systematic Review Screening Questions}
\label{appendix:a}

We provide the inclusion/exclusion used for the first screening in Table~\ref{tab:criteria}, and the QA Checklist used for the second screening in Table~\ref{tab:qachecklist}.

\begin{table}[htbp]
    \centering
    \renewcommand{\arraystretch}{1.8}
    \begin{tabular}{>{\raggedright\arraybackslash}m{0.2\textwidth} >{\raggedright\arraybackslash}m{0.75\textwidth}}
        \textbf{Type} & \textbf{Criteria} \\
        \hline
        \multirow{3}{*}{Inclusion} & 
        The paper needs to present or build on original social science theories or framework related to collaborative writing or dialogue \\
        & The paper is written in English \\
        & The paper focuses on English writing or English dialogue \\
        \hline
        \multirow{10}{*}{Exclusion} & 
        Not related to collaborative writing \\
        & The paper focuses on the pedagogical side of collaborative writing in education or collaborative learning \\
        & The paper focuses on cross-language collaborative writing, such as second language learners, writing in non-English languages, translation or interpretation \\
        & The paper focuses on story or fictional writing \\
        & The paper focuses on authorship or publishing ethics \\
        & The paper focuses on techniques for version control of documents \\
        & Proceedings titles \\
        & Unavailable for retrieval \\
        & Unpublished (i.e., “Early Access” papers on WoS) \\
        & Duplicated \\
        \hline
    \end{tabular}
    \vspace{10pt}
    \caption{Inclusion/exclusion criteria for the systematic review's screening process.}
    \label{tab:criteria}
\end{table}

\begin{table}[t]
    \centering
    \renewcommand{\arraystretch}{1.4}
    \begin{tabular*}{\columnwidth}{@{\extracolsep{\fill}}p{0.55\columnwidth}p{0.4\columnwidth}@{}}
        \textbf{Question} & \textbf{Scoring} \\
        \hline
        \textit{Does the paper introduce or build on original theories or frameworks related to collaborative writing?}
        &
        \begin{tabular}[t]{@{}l@{}}
            Yes (1 point)\\
            Partially (0.5 points)\\
            No (0 points)
        \end{tabular}
        \\
        \hline
    \end{tabular*}
    \vspace{10pt}
    \caption{QA checklist used in the systematic review.}
    \label{tab:qachecklist}
\end{table}

\clearpage
\section{The inductive synthesis process}
\label{appendix:b}
\begin{figure*}[!t]
  \centering
  \includegraphics[scale=0.13]{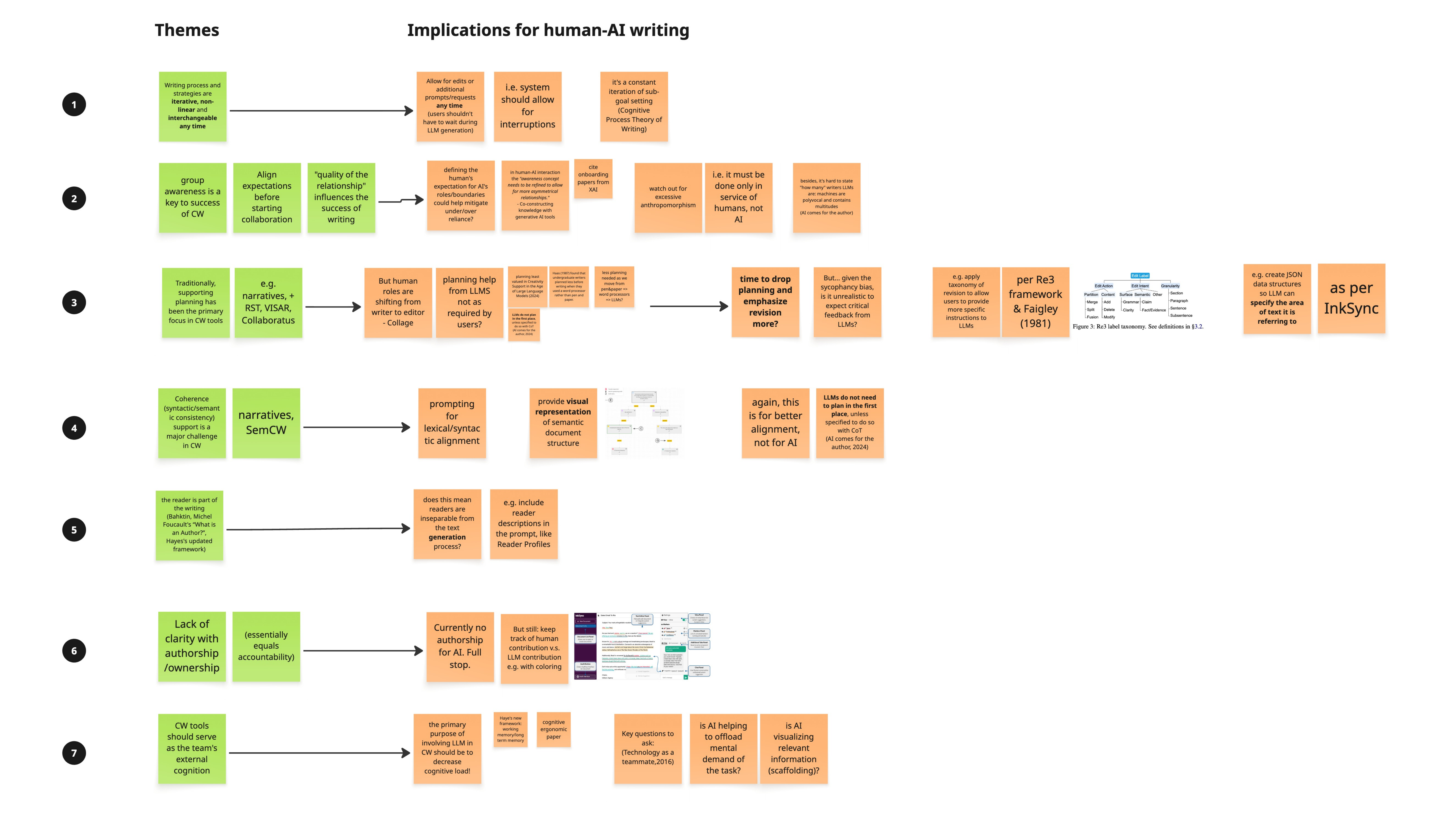}
  \caption{A screenshot of the inductive synthesis result on a shared visual workspace.}
  \Description{A diagram of granular types of revision, as proposed by Ruan et al.}
  \label{fig:miro-synthesis}
\end{figure*}

Figure~\ref{fig:miro-analysis} and Figure~\ref{fig:miro-synthesis}
show the ``raw'' inductive extraction and synthesis of the works considered. A high resolution version will be made available as auxiliary material with the final version of this manuscript.

\begin{figure*}[!b]
  \centering
  \includegraphics[scale=0.16, trim={0 101.7cm 0 0}, clip]{image/miro-analysis.pdf}
   \includegraphics[scale=0.16, trim={0 0 0 102cm}, clip]{image/miro-analysis.pdf}
  \caption{A screenshot of the extraction process on a shared visual workspace (split into two columns for easier viewing).}
  \Description{A diagram of granular types of revision, as proposed by Ruan et al.}
  \label{fig:miro-analysis}
\end{figure*}

\end{document}